\documentstyle[epsf,11pt]{article}
\newcommand{\bea}{\begin{eqnarray}}
\newcommand{\eea}{\end{eqnarray}}

\begin{document}
\begin{titlepage}

\thispagestyle{empty}

\vspace{0.2cm}

\title{{\normalsize \hfill HIP-2001-50/TH}\\
{\normalsize \hfill MRI-P-010904}
\\[20pt]
Radion and Higgs mixing at the LHC 
}
\author{Masud Chaichian$^{a,b}$, Anindya Datta$^{c}$, Katri Huitu$^a$ and
Zenghui Yu$^{a,b}$\\
$^a$Helsinki Institute of Physics\\
$^b$Department of Physics, University of Helsinki\\
P.O.Box 9, FIN-00014 Helsinki, Finland \\[4mm]
$^c$Harish-Chandra Research Institute
\\Chhatnag Road, Jhusi, Allahabad-211 019, India \\[7mm]}
\date{}
\maketitle

\vspace*{2truecm}

\begin{center}\begin{minipage}{5in}

\begin{center} ABSTRACT\end{center}
\baselineskip 0.2in

{We study the resonance production of radions and Higgs via
  gluon-gluon fusion in the Randall-Sundrum model with Higgs-curvature
  mixing at the LHC.  We find that radion can be detected both in
  mixed (with Higgs boson) and unmixed case if the radion vacuum
  expectation value $\Lambda_\phi$ is around 1 TeV. The
  $\Lambda_\phi \sim
  10$ TeV case is also promising for certain values of mixing
  parameters and radion masses.  The mixing can affect the
  production and decay of Higgs boson in a significant way.  Thus
  Higgs search strategies at the LHC may need refinements in case of
  radion-Higgs mixing in the Randall-Sundrum model.} \\

\vskip 5mm

%{~~~~PACS number(s): 13.65.+i, 13.88.+e, 14.65.-q, 14.80.Dq, 14.80.Gt}
\end{minipage}
\end{center}
\end{titlepage}

\eject
\rm
\baselineskip=0.25in

\begin{flushleft} {\bf I. Introduction} \end{flushleft}

\noindent
Recently proposed scenarios involving extra dimensions \cite{s1, s2}
provide an interesting possibility to probe the structure of the
space-time at TeV colliders. These models try to relate two
fundamental scales of physics, namely the Planck scale and the
electroweak scale.  All these models assume our world is $(4+n) + 1$
dimensional, where the extra $n$ space-like dimensions are curled up
with compactification radius smaller than the current experimental
reach.  The ADD \cite{s1} model requires relatively large
compactification radius ($\sim 1~mm$). In the following we will be
interested in the phenomenology of the model by Randall and Sundrum
(RS) \cite{s2}. RS model assumes our universe is $(4+1)$ dimensional.
Unlike the ADD, this scenario does not require a large
compactification radius for this extra compactified space-like
dimension. Moreover in RS model the radius of compactification is of
the order of Planck length and interestingly is a dynamical object.
It is connected to the vacuum expectation value of the dilaton field
arising due to compactification of full 5 dimensional theory to 4
dimensions.  Radion field is basically the exponential of this dilaton
field scaled by proper factors.  Goldberger and Wise have shown
\cite{s3,s3a} that one can write a potential for this radion field by
adding a scalar field to the bulk and dynamically generate the
VEV. It was also shown that without doing any fine tuning to the
parameters of the theory, this VEV can be of the order of TeV.  Radion
mass in the stabilised RS model comes out to be typically lighter than
the low-lying Kaluza-Klein modes of graviton \cite{s3, s4}. Thus
radion might be the first state, which is specific to the model and
accessible to the next generation TeV colliders.

The phenomenology of radions has been discussed in several works
\cite{s4,s5,s5a,s5b,new1,PSS}.  We will concentrate on the aspects of
Higgs-curvature mixing \cite{s6,new2} in this paper.  Mixing is due to
the following term in the action.
$$
S=-\xi \int d^4 x \sqrt{-g_{vis}} R(g_{vis}) H^\dagger H,
\eqno{(1.1)}
$$
The Ricci scalar $R\;(g_{vis})$ corresponds to the induced four
dimensional metric, $g_{vis}$, on the visible brane and $H$ is the
electroweak Higgs boson. This term will introduce mixing between
radion and Higgs in the RS model.  Since the Higgs search is one of
the main goals of the future collider experiments, the mixing of Higgs
with another particle is of major importance, if it will change the
Higgs production or decay patterns.  We will see in the following that
mixing of radion with Higgs will modify the Higgs and radion
phenomenology significantly.

The radion production via gluon-gluon fusion dominates the production
process of radion at LHC \cite{s5a}, and the effects of radion for Z
boson pair production have already been considered in \cite{PSS}.  The
effects of the Higgs-curvature mixing on radion production at LC were
also considered in \cite{our1}.  In the present paper we will discuss
the resonance production of radion and Higgs via gluon-gluon fusion
($pp\;(gg) \rightarrow h',\phi'$), which may probe a wide mass range
making it possible to study the  effects of curvature-Higgs
mixing at the LHC.\footnote{The states $h'$ and $\phi'$ are the 
physical ones after the mixing.}

In section 2, we will discuss the coupling of radion and Higgs to the
Standard Model (SM) fields, and in section 3 we discuss the decay
modes of Higgs and radion in the case of the curvature-Higgs mixing.
Section 3 will also contain the numerical results of radion and Higgs
production. We conclude in section 4. Some details of the
expressions are listed in the Appendix.

%\newpage

\begin{flushleft} {\bf II. Curvature-Higgs mixing in the
Randall-Sundrum model.} 
\end{flushleft}

\noindent
The action (1.1) leads to the
curvature-Higgs mixing Lagrangian \cite{s6,new2} given by
$$
{\cal L} = - 6\xi \Omega^2\left(\Box \ln \Omega 
+ (\nabla \ln  \Omega)^2\right)  H^{\dagger} H ,
\eqno {(2.1)}
$$
\noindent where 
$$\Omega=e^{-(\gamma/v) \phi(x)},\; 
\gamma=v/\Lambda_{\phi}.$$ 
Here $v$ is the Higgs VEV and $\Lambda_{\phi}$ is the radion VEV.

The interactions in (2.1) will induce the curvature-Higgs mixing, as 
discussed in the Appendix.
The couplings of  the physical radion and 
Higgs ($\phi^{'}$, $h^{'}$) to the SM gauge fields and fermions 
will be modified to
$$
\begin{array}{lll}
{\cal L}=-\frac{1}{\Lambda_{\phi}} (m_{ij} \bar{\psi}_i \psi_j - M_V^2
V_{A\mu}
V_{A}^{\mu}) \left[a_{34} \frac{\Lambda_{\phi}}{v} h^{'} 
+ a_{12} \phi^{'}\right],
\end{array}
\eqno{(2.2)}
$$
where $a_{12}= a+c/\gamma$ and
$a_{34}=d+b\gamma$, where $a,b,c,d$ are the mixing parameters given 
in the Appendix.
It is seen that the mixing changes significantly the couplings of 
Higgs and radion to the SM fields. 
For example, as pointed out 
in Ref. \cite{s6}, $a_{12}$ 
can be approximately zero in the conformal limit $m_h=0, \xi=1/6$ when
$\Lambda_{\phi}>>v$.  

The coupling of the radion to two Higgs bosons depends on the scalar
potential, $V(\phi)$ and mixing of radion and Higgs.  Neglecting the
radion self-coupling in $V(\phi)$, we can get the vertex of
$h^{'}h^{'}\phi^{'}$ as
$$
\begin{array}{lll}
V_{\phi^{'}h^{'}h^{'}}
&=&\frac{1}{\Lambda_{\phi}} \left(2 m_h^2 a d^2 h^{'2} \phi^{'} 
- a d^2 \phi^{'} \partial_{\mu} 
h^{'}
\partial ^{\mu} h^{'} (1-6\xi)  
+ 6\xi ad^2 (h^{'} \Box h^{'}) \phi^{'}\right.\\
&&\left.+ 4 m_h^2 b c d \phi^{'} h^{'2}-
2bcd h^{'} \partial_{\mu} \phi^{'} \partial^{\mu} h^{'} (1-6\xi)+
6bcd \xi h^{'} (\phi^{'}\Box h^{'}+h^{'} \Box \phi^{'})\right).
\end{array}
\eqno{(2.3)}
$$
Trace anomaly significantly modifies the radion and Higgs coupling to  
$gg$ and $\gamma \gamma$ \cite{s8}.  The effective vertices 
are given by
$$
V_{gg}=
\left[\frac{1}{\Lambda_{\phi}} (a b_3 - \frac{1}{2} a_{12} F_{1/2}(\tau_t))
\phi^{'}+\frac{1}{v} (\frac{v}{\Lambda_{\phi}} b b_3 - 1/2 a_{34} F_{1/2}
(\tau_t)) h^{'}\right] \frac{\alpha_{s}}{8\pi} G_{\mu\nu}^a G^{\mu\nu a}
\eqno{(2.4)}
$$
for radion and Higgs to gluons and
$$
\begin{array}{lll}
V_{\gamma\gamma}
&=&\left[\frac{1}{\Lambda_{\phi}} \{a (b_2+b_Y) - a_{12}
(F_1(\tau_W) +\frac{4}{3} F_{1/2}(\tau_t))\}
\phi^{'}+\right.\\
&&\left.\frac{1}{v} \{\frac{v}{\Lambda_{\phi}} b
(b_2+b_Y) - a_{34} (F_1 (\tau_W) + \frac{4}{3} F_{1/2}
(\tau_t))\} h^{'}\right] \frac{\alpha_{EM}}{8\pi} F_{\mu\nu} F^{\mu\nu}
\end{array}
\eqno{(2.5)}
$$
for radion or Higgs coupling to a pair of photons, where $b_3=7$ is
the QCD $\beta$-function coefficient and $b_2=19/6, b_Y=-41/6$ are the
$SU(2)\times U(1)_Y$ $\beta$-function coefficients in the SM. $F_1$
and $F_{1/2}$ are form factor from loop effects, which will be given
in detail in the Appendix. In each of these couplings the first term 
proportional to $b_3$ or $b_2 + b_Y$ are coming from the trace anomaly.
The rest are from the electroweak symmetry breaking.
 We can see from Eq.(2.4,5) that the
vertices Higgs-gluon-gluon and Higgs-photon-photon have new
contributions, which change the production and decay of the Higgs boson.

As seen in the Appendix, the mixing matrix of radion and Higgs is not
unitary.  Therefore it is not always straightforward, which particle
should be called Higgs and which should be called radion.  We
will always call $\phi^{'}$ radion and $h^{'}$ Higgs in the following
calculations.

\begin{flushleft} {\bf III. Radion and Higgs production and decay} 
\end{flushleft}

\noindent
The experimental groups at LHC have made thorough studies of the
possibilities to observe the Standard Model Higgs bosons at LHC.
The most straightforward detection modes, with the corresponding
Higgs mass ranges are (see e.g. \cite{ATLAS})
$$
\begin{array}{ll}
H\rightarrow\gamma\gamma, &100\;{\rm GeV}\;< \; m_H\;<\;150\;{\rm GeV},
\\
H\rightarrow WW, &150\;{\rm GeV}\;< \; m_H\;<\;190\;{\rm GeV},
\\
H\rightarrow ZZ, &190\;{\rm GeV}\;< \; m_H\;<\;\sim 700\;{\rm GeV}.
\end{array}
 \eqno{(3.1)}
$$
These are the decay modes of Higgs and radion that we will study in
this work.

\begin{figure}[t]
\leavevmode
\begin{center}
\mbox{\epsfxsize=6.truecm\epsfysize=7.2truecm\epsffile{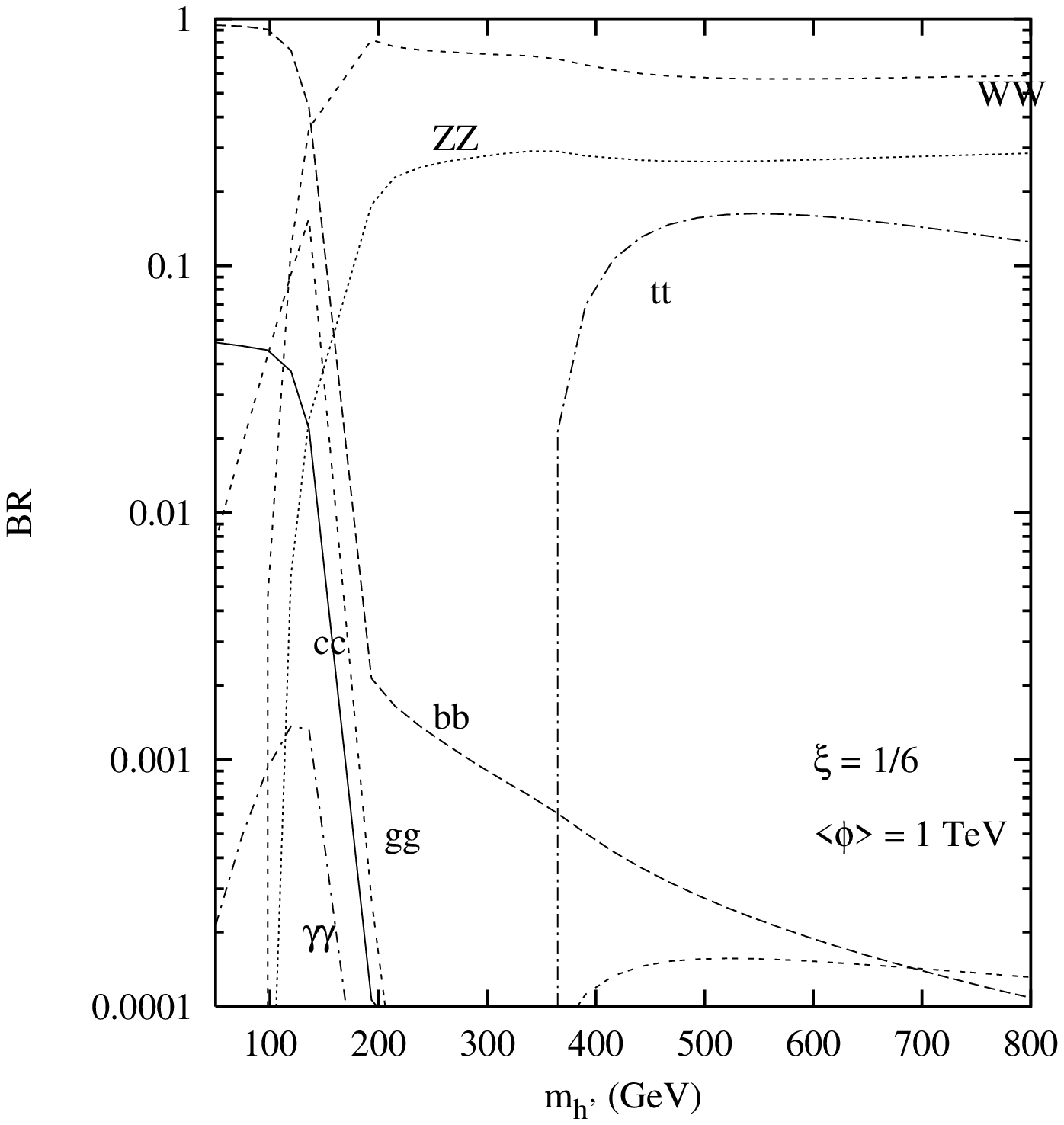}}
\mbox{\epsfxsize=6.truecm\epsfysize=7.2truecm\epsffile{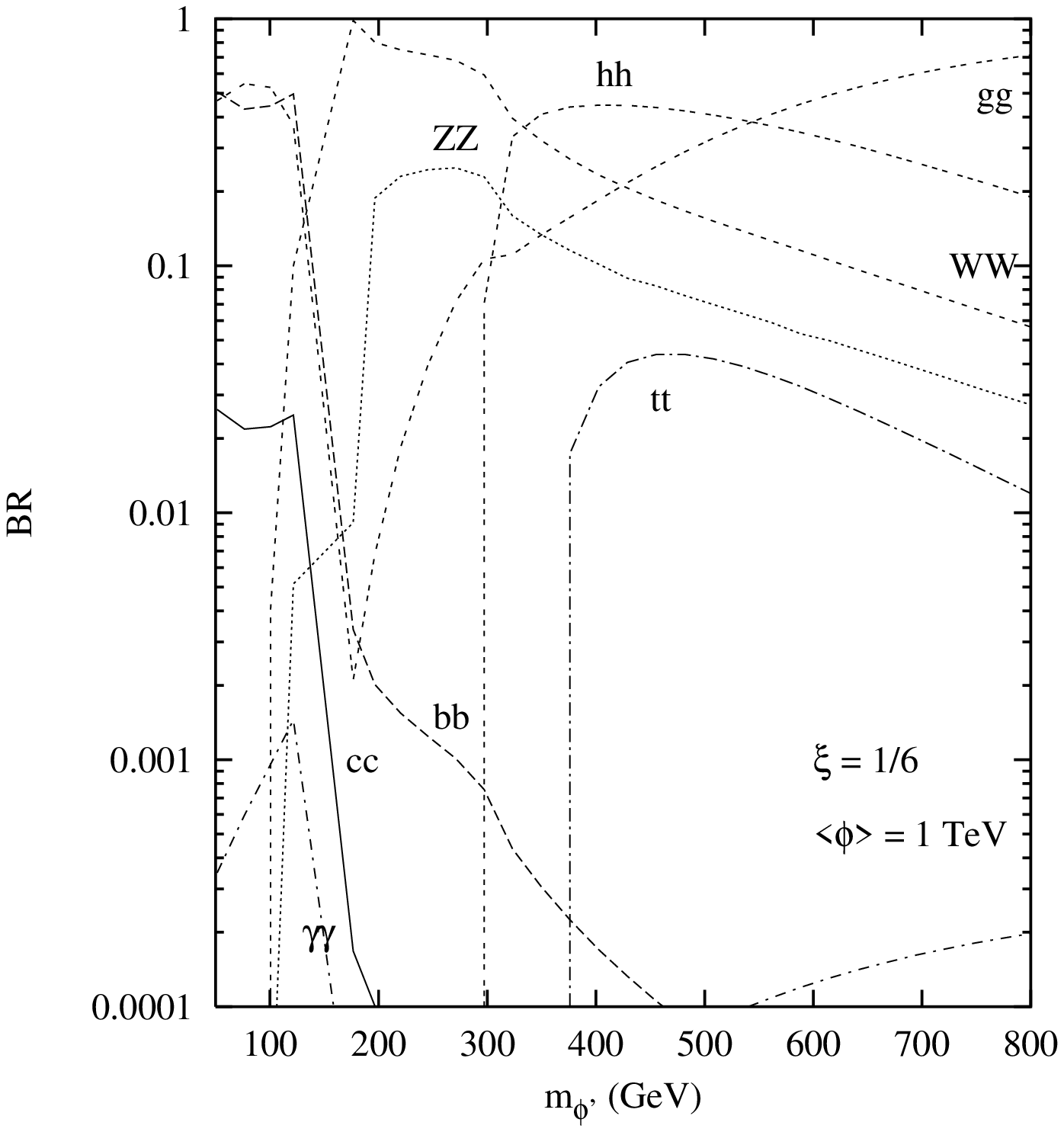}}
\end{center}
\caption{\label{fig1} Branching ratios of Higgs and radion decays
as functions of $m_{h'}$ and $m_{\Phi'}$, respectively.
We have used $\xi=1/6$, $\Lambda_\Phi=1$ TeV, and
$m_h=150$ GeV in the plots.
  }
\end{figure}
Because of the mixing the decay patterns of Higgs and radion will
change.  In Figure 1 we present the decay branching ratios of Higgs
and radion in the mixed case, with the mixing parameter $\xi=1/6$,
$\Lambda_\Phi=1$ TeV, and $m_h=150$ GeV.  From Fig. 1 it is evident
that when $h^{'}$ is heavier than 180 GeV the decay of Higgs to WW and
ZZ will dominate.  For radions heavier than twice the physical Higgs
mass, the decay to Higgs and for larger masses to gluons will be
dominant.  If the radion VEV is increased, the branching ratios do not
change significantly.  Compared to the unmixed case, the branching
ratios of heavy radion are changed.  In the unmixed case, also for
heavy radion, the dominant decay modes are the weak gauge bosons, and
the branching ratio to gluons is at the percentage level
\cite{s5a,our1}. There are few other interesting points which we want
to point out. The $h'$ branching ratio to gluons has a sharp dip
around $m_{h'} = 250$ GeV. This can be explained by the structure of
the $h'gg$ coupling. This coupling has two terms. Second term is
complex when $h'$ mass is greater than $2m_t$. (The imaginary part
does not bother us as the first term is real, so for the cancellation
real part of the second term is more important). For $\Lambda_\phi = 1$
TeV, $\xi = 0.167$, around $m_{h'} = 250$ GeV, there is a
cancellation between these two terms. This drives the $h'$ width to
two gluons to zero around this mass region. We will see that this will also
affect the $h'$ signals.  When one changes
$\Lambda_\phi$ to $10$ TeV, second term changes very little but the first
term is modified (its magnitude is reduced) and thus the cancellation is
not so severe in this case. Again when  the sign of $\xi$ is changed,
(i.e $\xi= -0.167$) one can easily check that the first term simply 
changes its
sign, while the second term remains almost unchanged. Thus the accidental
cancellation between two terms in $gg$ coupling shows up
only in $\xi >0$ case. There is no such cancellation in $\phi ' gg$
coupling.

\begin{figure}[t]
\leavevmode
\begin{center}
\mbox{\epsfxsize=6.truecm\epsfysize=6.truecm\epsffile{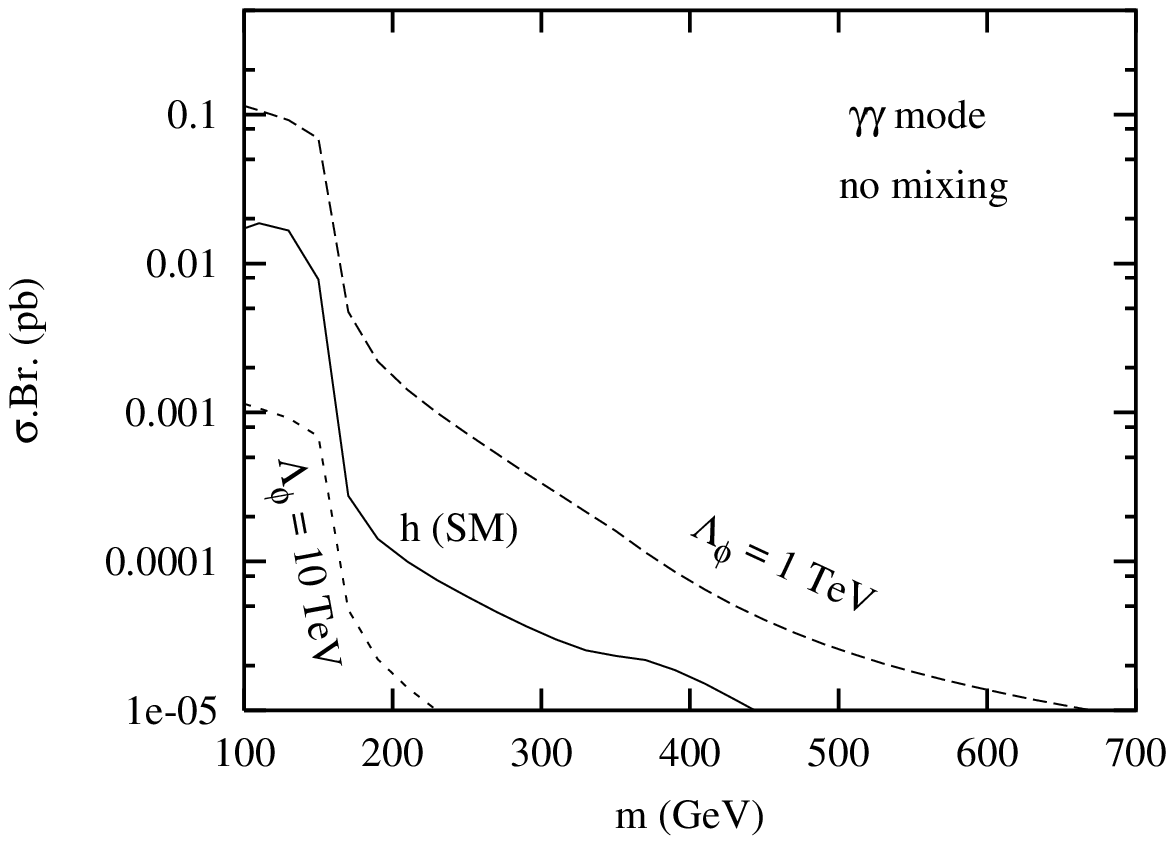}}
\mbox{\epsfxsize=6.truecm\epsfysize=6.truecm\epsffile{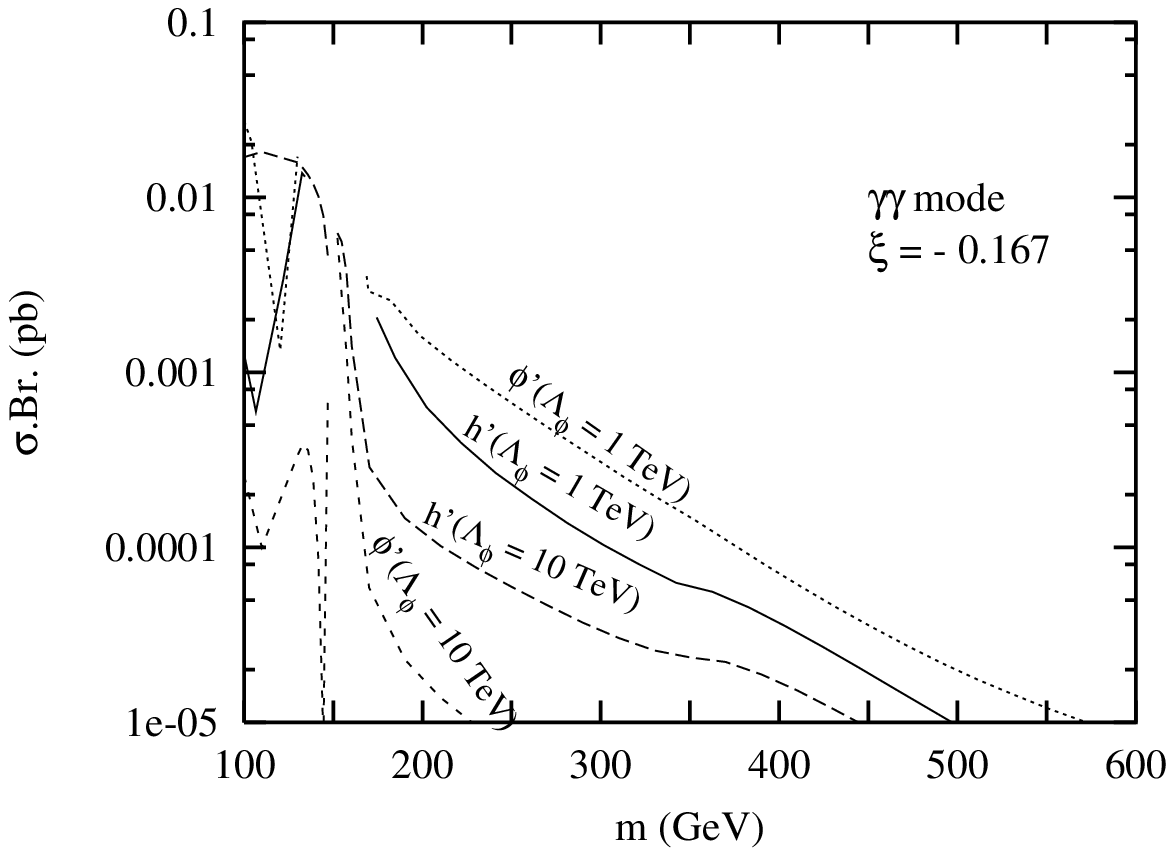}}
\mbox{\epsfxsize=6.truecm\epsfysize=6.truecm\epsffile{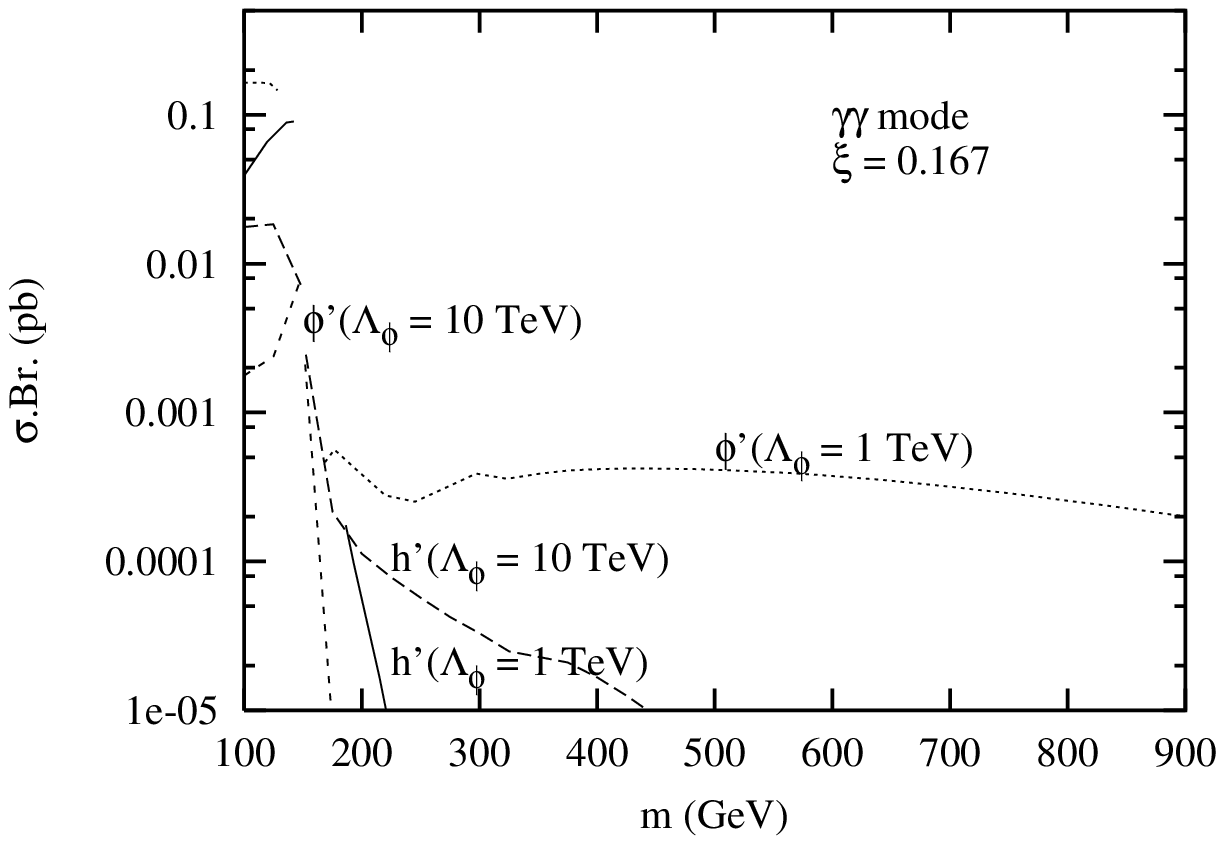}}
\end{center}
\caption{\label{fig2} The cross-section of $ pp (g g) \rightarrow
\phi^{'} (h^{'}) \rightarrow \gamma \gamma$ as a function
of $m_{\phi^{'}}$ ($m_{h^{'}}$) with $\Lambda_{\phi}=1$ TeV
and $10$ TeV, $m_h$=150 GeV,
(a) $\xi=0$, (b) $\xi=-1/6$, and (c) $\xi=1/6$.}
\end{figure}

\begin{figure}[t]
\leavevmode
\begin{center}
\mbox{\epsfxsize=6.truecm\epsfysize=6.truecm\epsffile{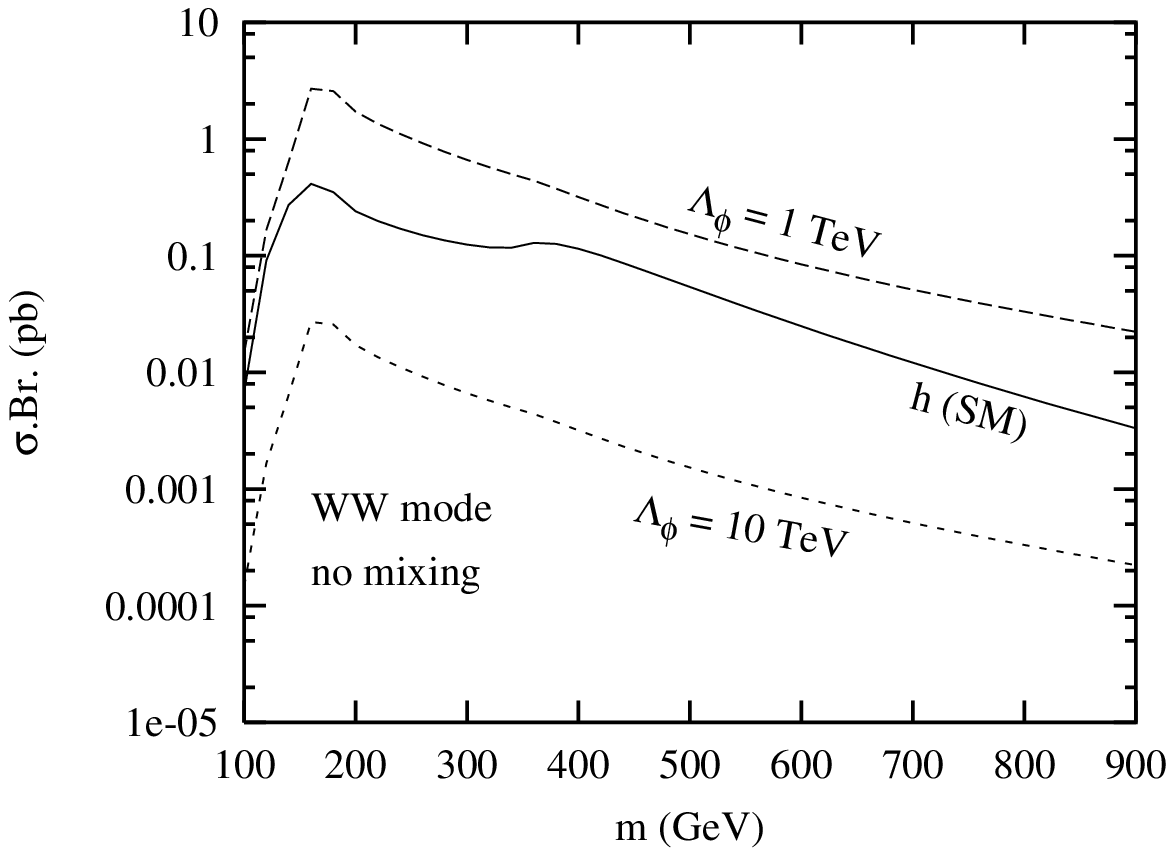}}
\mbox{\epsfxsize=6.truecm\epsfysize=6.truecm\epsffile{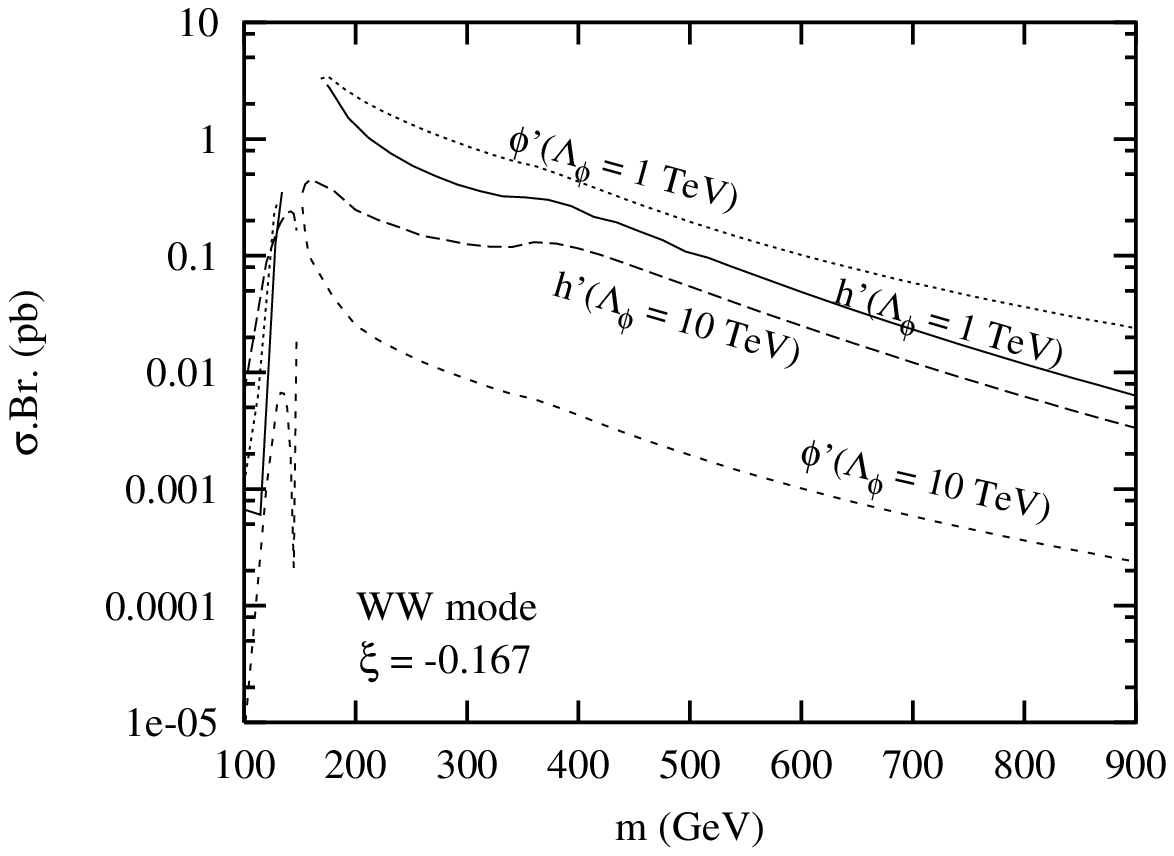}}
\mbox{\epsfxsize=6.truecm\epsfysize=6.truecm\epsffile{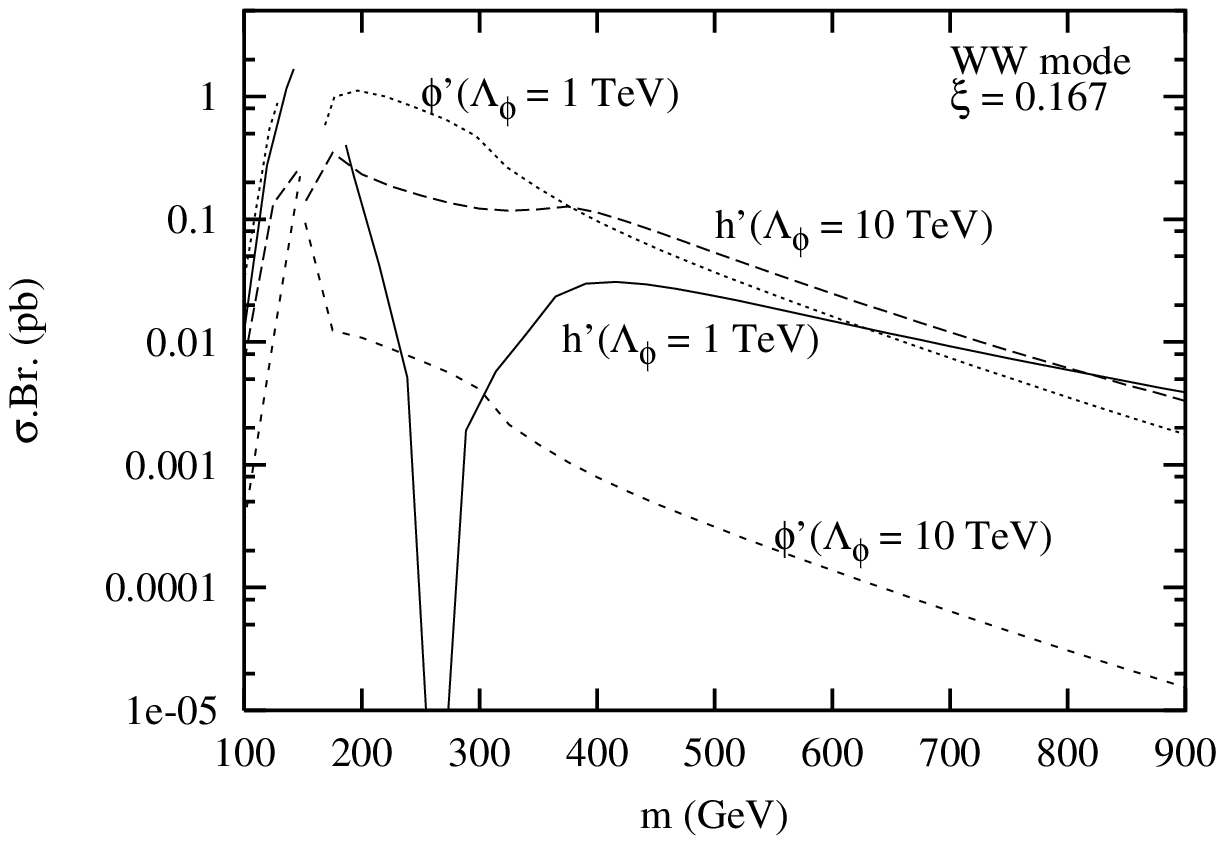}}
\end{center}
\caption{\label{fig3} The cross-section of $pp (g g) \rightarrow
\phi^{'} (h^{'}) \rightarrow W^+W^- \rightarrow l^+ l^- \nu 
\bar \nu \;(l \equiv e,\mu)$ as a function
of $m_{\phi^{'}}$ ($m_{h^{'}}$) with $\Lambda_{\phi}=1$ TeV
and $10$ TeV, $m_h$=150 GeV,
(a) $\xi=0$, (b) $\xi=-1/6$, and (c) $\xi=1/6$.}
\end{figure}

Next we will discuss the production cross-sections of Higgs and radion
in gluon-gluon fusion, multiplied by the branching ratios to
$\gamma\gamma$, $ZZ$ and $WW$ decay modes ($ZZ$ and $WW$
cross-sections will be further multiplied by the branching ratios $Z
\rightarrow l^+l^-$ and $W \rightarrow l \nu_l$; where $l \equiv
e,\mu$).  Gluon-gluon fusion is the dominant production process for
the Higgs and production cross-section is further enhanced by the
trace anomaly in radion production.  In all the figures that we'll
present in the following, the c.m. energy of the LHC is assumed to be
14 TeV. We have used $m_h$ = 150 GeV in all our following analysis.

In Figure 2, we show the cross-section of $pp (gg) \rightarrow
\phi^{'} (h^{'}) \rightarrow \gamma \gamma$ as a function of
$m_{\phi^{'}}$ ($m_{h^{'}}$). Fig. 2 (a) corresponds to the case
$\xi=0$.  For the Higgs production the cross-section is same as in the
Standard Model.  When calculating the proton-proton cross-section from the
parton level (in this case and in the following), we have used CTEQ4L
parton distribution functions \cite{cteq} with factrorization scale
set at $m_{\phi',h'}.$ As seen from the Figure, the radion cross
section depends strongly on the radion VEV, $\Lambda_{\phi}$.  The
cross-section for $\Lambda_{\phi}=1$ TeV is larger than the Higgs
cross-section because of the anomaly, extending the detectability of
the mode beyond $m_{\phi'}=160$ GeV, but the suppression by the radion
VEV is evident, when $\Lambda_{\phi}=10$ TeV.  In Fig. 2 (b) and (c)
we consider the $\xi \neq 0$ case.  Higgs cross-section (in 2(b) and
2(c)) also depends on $\Lambda_{\phi}$ because of the mixing.
Furthermore, it is clear that the absolute value, as well as the
sign of the mixing parameter are crucial for the cross-section.  In
(b) we set $\xi=-1/6$, and in (c) $\xi=1/6$.  The Higgs curvature
mixing changes the situation dramatically.  For the positive mixing,
the $\gamma\gamma$ cross-section is increased for the physical Higgs mass
$m_{h'}<150$ GeV, while for the negative mixing the cross-
section decreases.  For $\xi=1/6$, in Fig. 2 (c), if both
scalars are lighter than $\sim 145$ GeV, they can be detected with
enhanced cross-sections if $\Lambda_\Phi =1$ TeV.  If
$\Lambda_\Phi=10$ TeV, for a very small range close to 145 GeV two
scalars may be detectable.  In the case of negative mixing, in Fig. 2 b,
the cross-sections for $\Lambda_\Phi =1$ TeV are decreased.  Two
scalars are visible if they are both rather degenerate, with masses
around 125 GeV.  If $\Lambda_\Phi=10$ TeV, one scalar can be detected
if it is lighter than 150 GeV.  The discontinuities in the scalar
masses in the plots are due to the discontinuity in the radion-Higgs
mass matrix elements, as seen in the Appendix. We want to point out
another gross feature of these plots.  For scalar masses less than
$2m_W$, $\gamma \gamma$ cross-section remains almost constant (for
no-mixing case) or changes slowly with mass (for $\xi \neq 0$). As
soon as the $2W$ decay mode is open, $\gamma \gamma$ branching ratio
falls off pretty fast for all of the above cases, decreasing the intensity
of $\gamma \gamma$ signal beyond this mass range. The sudden dip and
jump of the $pp \rightarrow \phi' \rightarrow \gamma \gamma, ZZ, WW$
cross-section for $\Lambda_\Phi=10$ TeV , $\xi= - 1/6$ around
$m_{\phi'} = 150$ GeV again can be accounted by the variation of
$a_{12}$ in $\phi'gg$ coupling and also by the choice of the value of
the Higgs mass parameter ($m_h$) in our analysis.

\begin{figure}[t]
\leavevmode
\begin{center}
\mbox{\epsfxsize=6.truecm\epsfysize=6.truecm\epsffile{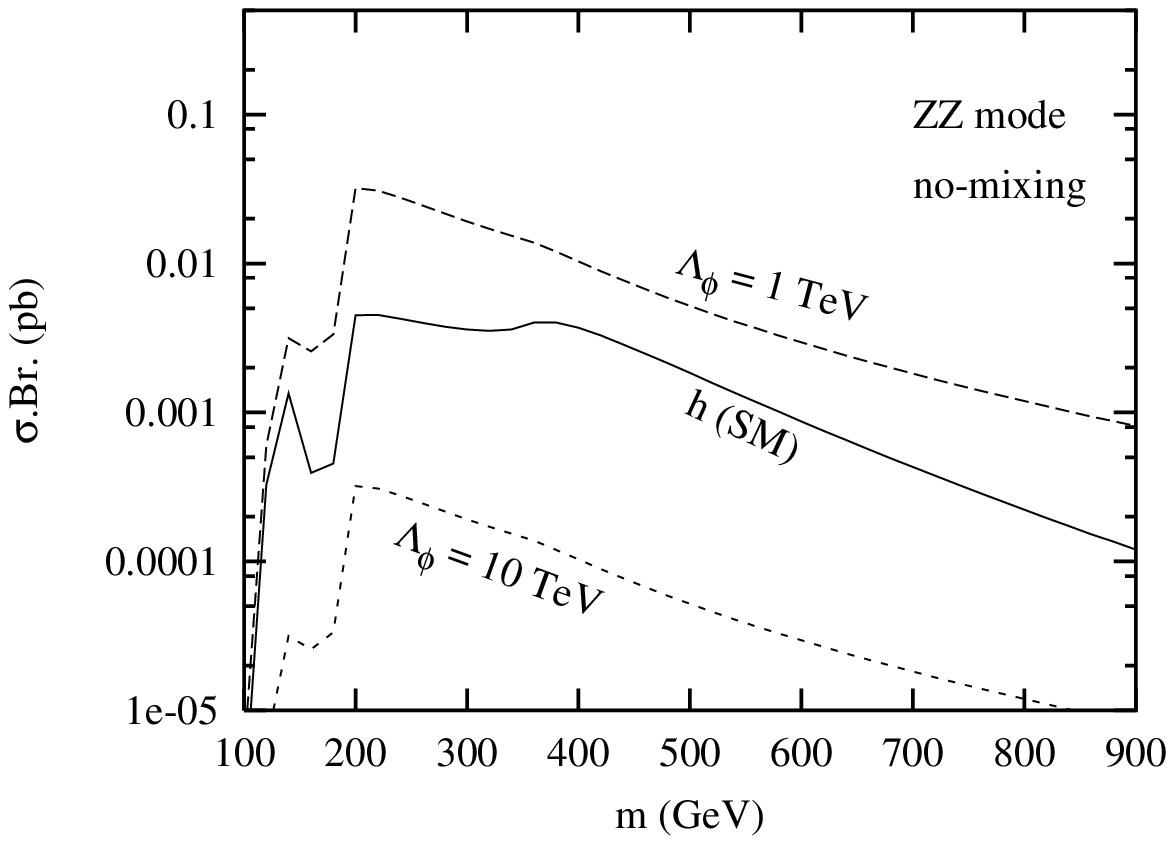}}
\mbox{\epsfxsize=6.truecm\epsfysize=6.truecm\epsffile{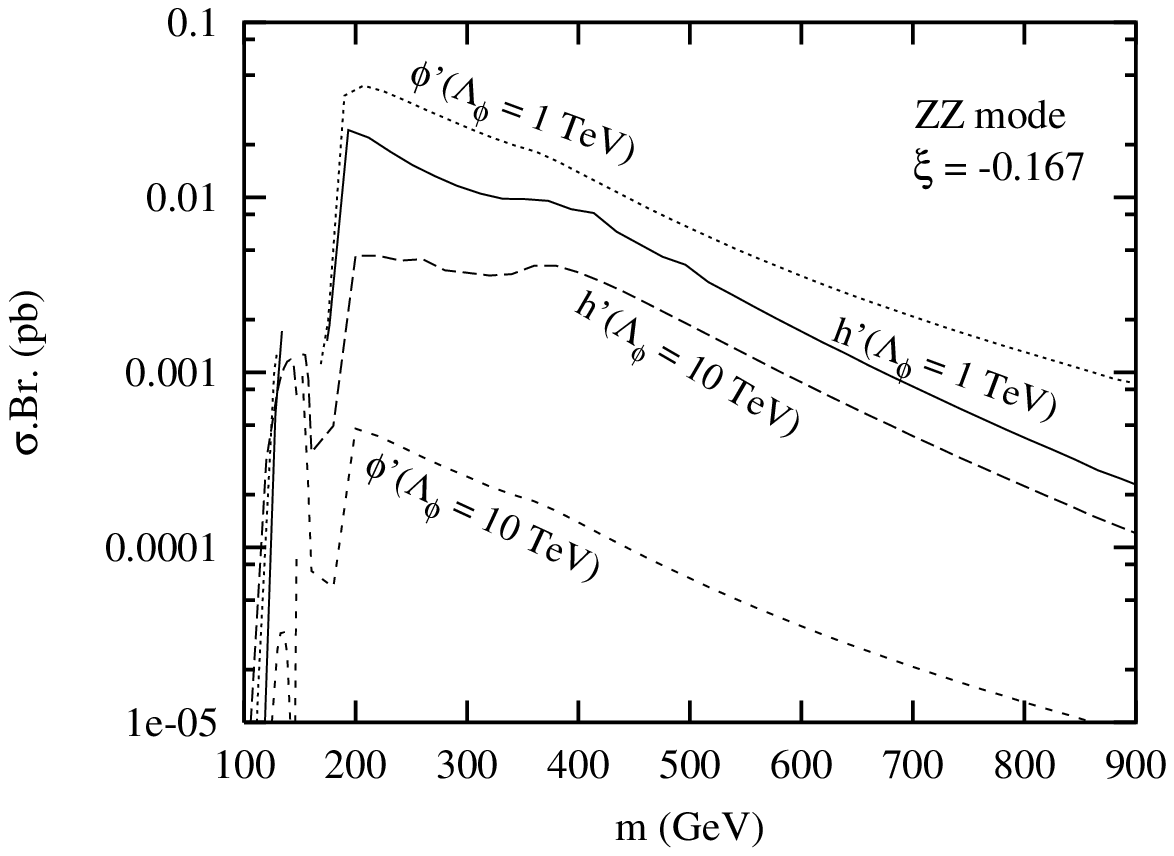}}
\mbox{\epsfxsize=6.truecm\epsfysize=6.truecm\epsffile{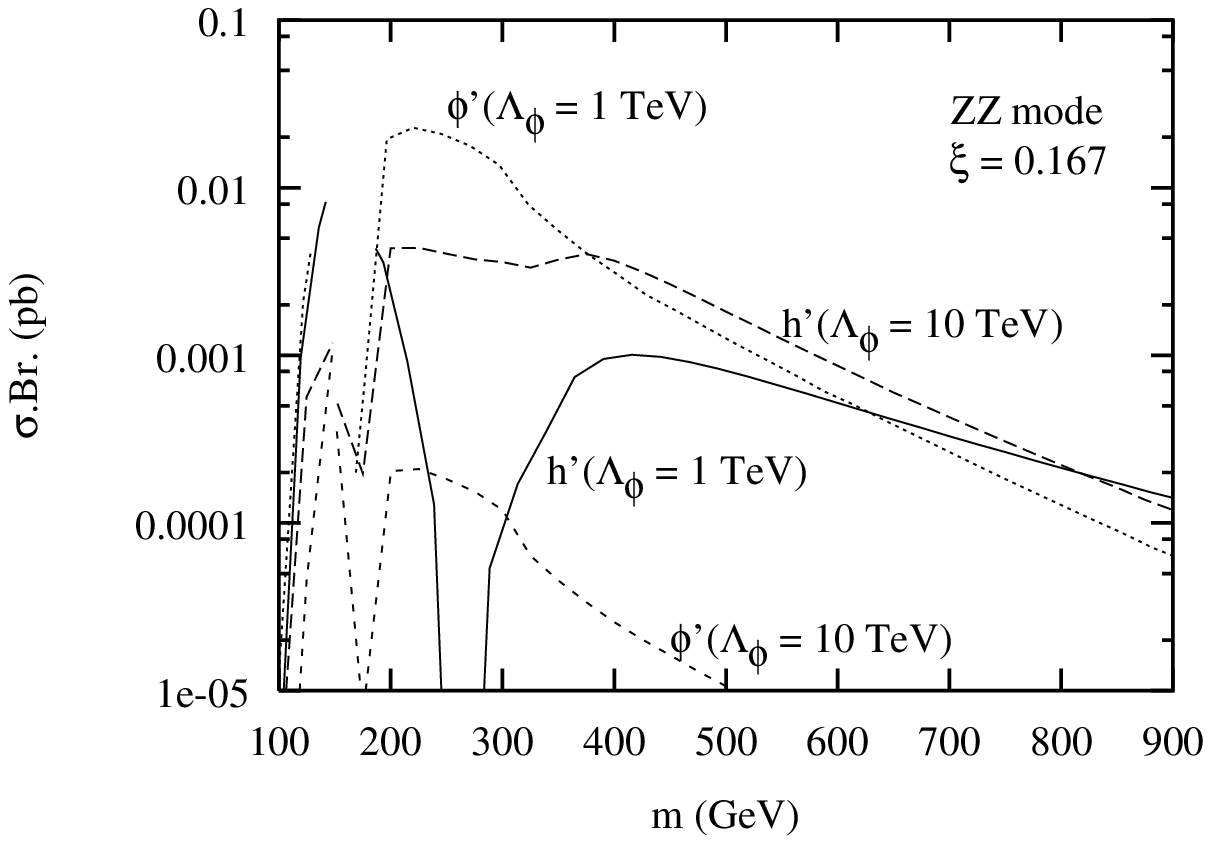}}
\end{center}
\caption{\label{fig4} The cross-section of $ pp (g g) \rightarrow
\phi^{'} (h^{'}) \rightarrow ZZ \rightarrow l^+ l^- l^+ l^- 
(l \equiv e,\mu)$ as a function
of $m_{\phi^{'}}$ ($m_{h^{'}})$ with $\Lambda_{\phi}=1$ TeV   
and $10$ TeV, $m_h$=150 GeV,
(a) $\xi=0$, (b) $\xi=-1/6$, and (c) $\xi=1/6$.}
\end{figure}

In Figure 3 we present the corresponding plots for the process $pp
(gg) \rightarrow \phi^{'} (h^{'})\rightarrow W^+W^- \rightarrow l^+
l^- \nu \bar \nu \;(l \equiv e,\mu)$, which becomes important in the
Standard Model case when the Higgs mass is above approximately 150
GeV.  In Fig. 3 (a), we set $\xi=0$.  While the $\gamma\gamma$ mode is
useful for radion detection up to 200 GeV when $\Lambda_\phi =1$ TeV,
the $WW$ mode is observable up to around 400 GeV.  For the $WW$ mode
with $\Lambda_\phi=1$ TeV there is a clear increase in the Higgs
cross-section, if the mixing parameter is negative, and two scalars
should be observable if they are lighter than 400 GeV.  If
$\Lambda_\phi=10$ TeV, the Higgs looks similar to the Standard Model
Higgs, while radion is unobservable.  If the mixing parameter is
positive, only one scalar is detectable over most of the mass range,
nearly up to 300 GeV, for $\Lambda_\phi=1$ TeV.  Only for light
masses, below 140 GeV, two scalars could be detected.  For
$\Lambda_\phi=10$ TeV, one scalar may be detectable around 200 GeV.
Again the sudden dip in the $WW$ (and also in $ZZ$, which we discuss
in the following paragraph) production cross-section (via resonance
$h'$ production) around $m_{h'}$ = 250 GeV for $\Lambda_\phi=1$ TeV,
$\xi=1/6$, can be explained by the vanishing $h'gg$ coupling around
this mass range.

In Figure 4 we plot the cross-section for $pp (gg) \rightarrow
\phi^{'} (h^{'}) \rightarrow ZZ \rightarrow l^+ l^- l^+ l^- (l \equiv
e,\mu)$, with (a), (b), and (c) corresponding to similar sets of
parameters than in the Figures 2 and 3.  In the Standard Model this
process provides the golden signal of Higgs production, four leptons
with no missing energy.  The $H\rightarrow ZZ\rightarrow 4l$ is the
best signal for Higgs in the range 180 GeV$<m_H<\sim 700$ GeV
\cite{ATLAS}.  For $\Lambda_\phi=1$ TeV, there is again obvious
increase for the Higgs cross-section if $\xi=-1/6$, and the mass range
for detecting two scalars increases nearly upto to 800 GeV. For 10 TeV the
effects are minor when compared to the Standard Model, except that
there are small ranges below 150 Gev, where two scalars can be
detected, both for positive and negative mixing parameter.

%-----------------------------------------------------------------------
\begin{figure}[h]
\leavevmode
\begin{center}
\mbox{\epsfxsize=6.truecm\epsfysize=6.truecm\epsffile{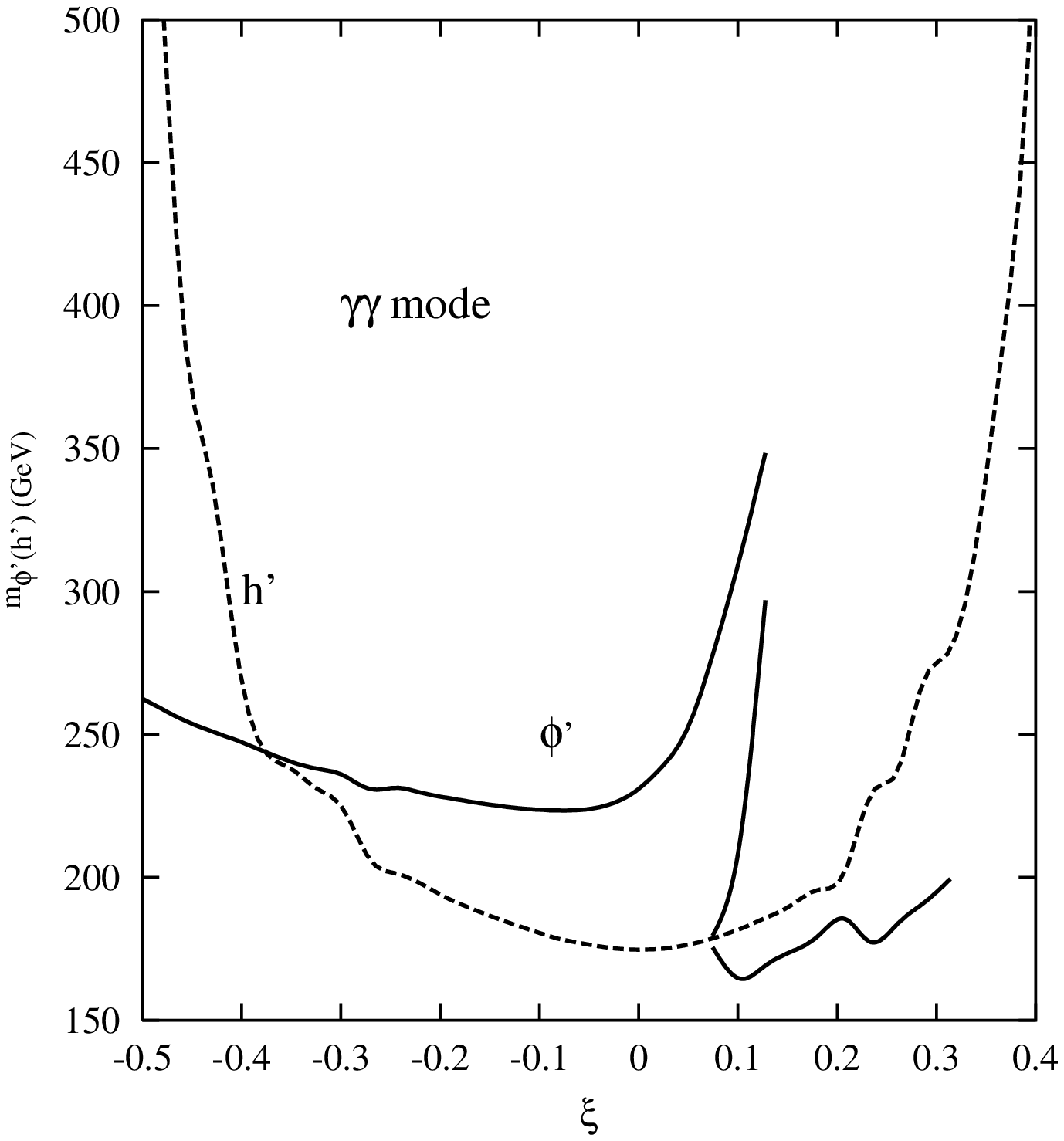}}
\mbox{\epsfxsize=6.truecm\epsfysize=6.truecm\epsffile{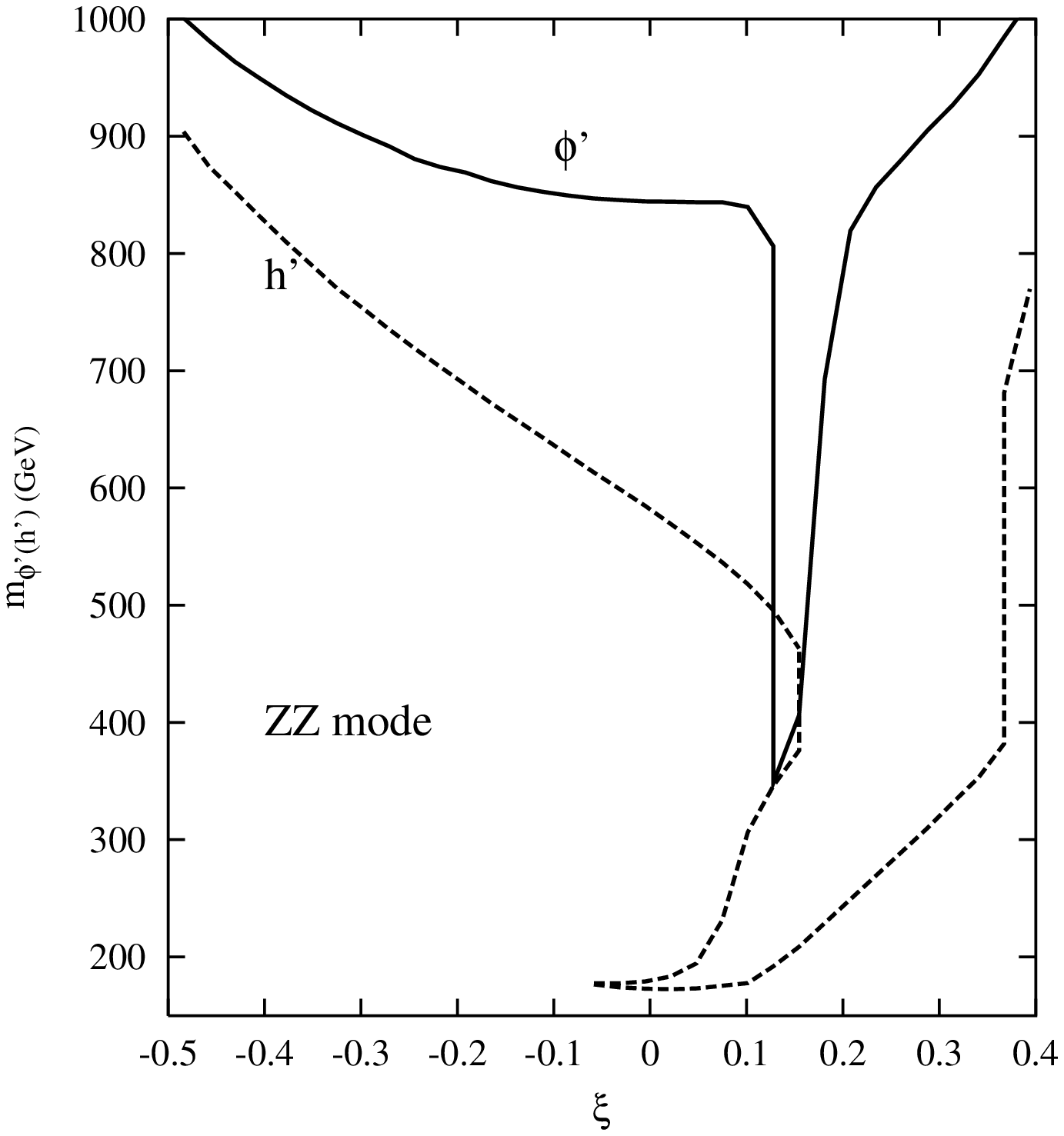}}
\end{center}
\caption{\label{fig5} Contours of 100 (a) $\gamma \gamma$  and 
(b) $ZZ$ events from
$\phi'$ (solid line) and $h'$ (dashed line) production and decay at 
the LHC in the ($\xi , m_{h'(\phi')}$) plane. 
We have assumed $\Lambda_{\phi}$ = 1 TeV, $m_h \;(m_\phi) = 150$ GeV and 
an integrated luminosity of 100 $fb^{-1}$.
}
\end{figure}
%--------------------------------------------------------------------------

Till now we have considered some specific values of the mixing
parameter $\xi$. Before we conclude, we discuss how the
$\gamma\gamma$ and $ZZ$ event rates vary with this parameter. From the
consideration of perturbative unitarity, $|\xi| > 3$ is ruled out
\cite{unitary}.  For $\Lambda_{\phi}$ = 1 TeV, $\xi$ can vary from
$-0.75$ to $0.56$, as one finds from the Appendix A.  The role of $\xi$ is
crucial in determining the physical masses of the scalars as well as
the couplings of them to the SM fields.  The effect of this parameter
on the event rates is twofold.  In Fig. \ref{fig5}, we have plotted the
contours of 100 $\gamma \gamma$ and $ZZ \;(l^+ l^- l^+ l^-)$ events from
$\phi'$ and $h'$ production and decay at the LHC in ($\xi, m_{h'(\phi')}$)
--plane.
For this calculation we have assumed $\Lambda_{\phi}$ = 1 TeV , $m_h\;
(m_\phi) = 150$ GeV and an integrated luminosity of 100 $fb^{-1}$.

Sudden jumps in the event contours around specific values of $\xi$ are
due to the discontinuities in the physical masses and sharp
maxima/minima of the relevant couplings, which we have pointed out
earlier.  Especially, at around $\xi\sim 0.3$, the
$\phi'-\gamma-\gamma$ coupling strongly reduces making the
corresponding mass reach too low to be shown in the Figure \ref{fig5}.
Apart from these irregularities, both $\phi'$ and $h'$ event rates
significantly increase with the absolute value of the mixing parameter
for the $ZZ \rightarrow l^+ l^- l^+ l^-$ channel, and $h'$ rates for
the $\gamma\gamma$ channel.  The $h'$ event rate is more sensitive to
the mixing parameter which is evident from both the channels.  We have
checked that physical masses are nearly symmetric with respect to the
positive and negative values of $\xi$.  The $h'$ event rate via
$\gamma \gamma$ channel is almost symmetric in positive and negative
values of $\xi$.  Almost for all negative values of $\xi$, mass reach
for $\phi' \rightarrow \gamma \gamma$ channel is better than for
$h'\rightarrow \gamma \gamma $ channel.  In the $ZZ$ channel, almost for any
$\xi$, the mass reach is better for $\phi'$, and the $\phi'$ event
rate is symmetric about $\xi = 0$ apart from one sharp dip around $\xi
\simeq 0.1$. This is due to the sudden dip in the $\phi'-g-g$
coupling. We have not presented the corresponding plots for the $WW$
channels.  Production mechanism for both the $WW$ and $ZZ$ channels
are the same.  The $\phi', h' \rightarrow WW \rightarrow l^+ \nu l^-
\bar \nu$ effective branching ratio is almost an order of magnitude
greater than that of $\phi', h' \rightarrow ZZ \rightarrow l^+ l^- l^+
l^-$ channel.  Therefore, it is evident that the mass reach of the
$WW$ channel is better than that of the $ZZ$ channel for a particular
value of $\xi$. The dependence of the mass reach on $\xi$ for the $WW$
channel is the same as for the $ZZ$ channel.

\begin{flushleft} {\bf IV. Conclusion} \end{flushleft}

\noindent
We have studied the Higgs and radion production via gluon-gluon
fusion in the RS model with curvature-Higgs mixing.  Our results show
that radion and Higgs production from $gg$ collision will be very
different in mixed and unmixed cases.  Thus the detection of Higgs or
radion at the LHC may reveal the mixing strength, including sign, in
the model.

The decay modes of radion and Higgs in the mixing case will be quite
different from unmixed case. Especially, the abnormal coupling of
radion to gauge bosons can effect Higgs decay through mixing, thus
modifying the Higgs decay strongly.  When two Higgs like scalars are
seen, the different decay branching ratios, when compared to the
minimal supersymmetric standard model (MSSM), will help to distinguish
between MSSM and the RS model.

\begin{flushleft} {\bf Acknowledgement} \end{flushleft}

\noindent
The authors thank the Academy of Finland
(project number 163394 and 48787) for financial support.
Z.-H. Yu thanks the World Laboratory, Lausanne, for the scholarship. A. Datta 
acknowledges the hospitality of Helsinki Institute of Physics where this
work has been done.
 
\vspace*{2cm}
% \newpage
\begin{center} {\bf Appendix} \end{center}
\par
{\bf A. The curvature-Higgs mixing:}

After shifting $\phi\rightarrow \phi +
\Lambda_{\phi}$ in Eqn. (2.1), the Lagrangian containing bilinear 
terms of radion and Higgs is obtained as
$$
{\cal L} = - \frac{1}{2}  \phi [(1-6\xi \gamma^2) \Box + m_{\phi} ^2 ]
\phi
- \frac{1}{2} h (\Box + m_{h}^2 ) h - \frac{6 \xi v}{\Lambda_{\phi}}
\phi \Box h.
\eqno {(A.1)}
$$
Here $m_{\phi}$ is a mass parameter for $\phi$.

After diagonalisation, the fields should be redefined as
$$
\phi= a \phi^{'} + b h^{'},
\eqno {(A.2)}
$$
$$
h= c \phi^{'} + d h^{'},
\eqno {(A.3)}
$$
where
$ a=\cos \theta /Z$; $b=-\sin \theta/Z$;
$c=\sin \theta-6\xi \gamma/Z \cos \theta$ and
$d=\cos \theta +6 \xi \gamma /Z \sin \theta$,
with
$Z^2=1-6\xi \gamma^2 (1+6\xi)$ and
the mixing angle $\theta$ is given by
$$ \tan 2 \theta = 12 \xi \gamma Z 
\frac{m_h^2}{m_h^2(Z^2-36\xi^2\gamma^2)-
m_{\phi}^2}. 
\eqno {(A.4)}
$$   
Our results agree with
those in Ref. \cite{s6} (with $\xi \gamma <<1$) and in Ref. \cite{new2}.
{}From Eq. (A.3-4), we see clearly the
constraints $-(1+\sqrt{1+4/\gamma^2})/12 \le \xi \le
(\sqrt{1+4/\gamma^2}-1)/12$, just as in Ref. \cite{new2}.

The new fields $\phi^{'}$ and $h^{'}$ are mass eigenstates with masses
$$
m_{\phi^{'}}^2= c^2 m_h^2 + a^2 m_{\phi} ^2,
\eqno {(A.5)}
$$
$$
m_{h^{'}}^2= d^2 m_h^2 + b^2 m_{\phi} ^2.
\eqno {(A.6)}
$$
\par
The interaction Lagrangian of $\phi$ and $h$ with fermions and
massive gauge bosons,
$$
\begin{array}{lll}
{\cal L} =-\frac{1}{v} (m_{ij} \bar{\psi}_i \psi_j - M_V^2 V_{A\mu}
V_{A}^{\mu}) \left[h+ \frac{v}{\Lambda_{\phi}} \phi\right],
\end{array}
\eqno{(A.7)} 
$$
can be transformed to the coupling of mass eigenstates $\phi^{'}$ and
$h^{'}$ to fermions and massive gauge bosons as
$$
\begin{array}{lll}
{\cal L}=-\frac{1}{\Lambda_{\phi}} (m_{ij} \bar{\psi}_i \psi_j - M_V^2
V_{A\mu}
V_{A}^{\mu}) [a_{34} \frac{\Lambda_{\phi}}{v} h^{'} + a_{12} \phi^{'}],
\end{array}
\eqno{(A.8)} 
$$
where $a_{12}= a+c/\gamma$ and
$a_{34}=d+b\gamma$.
The coefficients $a_{12}$ and $a_{34}$ give directly the strength of
the corresponding interaction when compared to the case with no
mixing.

\noindent
{\bf B. Form factors}
\par
The form factors $F_{1/2} (\tau_t)$ and $F_{1} (\tau_W)$
can be defined as \cite{s6,s11}
$$
\begin{array}{lll}
F_{1/2} (\tau) =-2 \tau [1+(1-\tau) f(\tau)]
\end{array}
\eqno{(B.1)}
$$  
and
$$  
\begin{array}{lll}
F_{1} (\tau) =2 + 3 \tau + 3 \tau (2-\tau) f(\tau),
\end{array}
\eqno{(B.2)}
$$
where $\tau_t=4m_t^2/q^2$, $\tau_W=4m_W^2/q^2$ and
$$
\begin{array}{lll}
f(\tau) &=&[sin^{-1} (1/\sqrt{\tau})]^2,  \ \ \ \tau \ge 1, \\
&& -1/4 [Log (\eta_{+}/\eta_{-}) - i \pi]^2, \ \tau <1,
\end{array}
\eqno{(B.3)}
$$
with $\eta_{\pm}=1\pm \sqrt{1-\tau}$.

\end{document}